\documentclass[multphys,vecphys]{svmult}

\usepackage{makeidx}         
\usepackage{graphicx}        
\usepackage{multicol}        
\usepackage[bottom]{footmisc}

\makeindex             

\begin{document}

\title*{Spectral and variability properties of LS 5039 from radio
to very high-energy gamma-rays}
\titlerunning{Spectral and variability properties of LS 5039} 

\author{V. Bosch-Ramon\inst{1}\and
J. M. Paredes\inst{1}\and
G. E. Romero\inst{2,3}}
\institute{Departament d'Astronomia i Meteorologia, Universitat de Barcelona, 
Av. Diagonal 647, E-08028 Barcelona, Catalonia (Spain) \texttt{vbosch@am.ub.es,jmparedes@ub.edu}
\and
Instituto Argentino de Radioastronomía, C.C.5, (1894) Villa Elisa, Buenos Aires (Argentina) \texttt{romero@iar.unlp.edu.ar}
\and
Facultad de Ciencias Astronómicas y Geofísicas, UNLP, Paseo del Bosque, 1900 La Plata (Argentina)
}
%
%
\maketitle

\section{Introduction}
\label{sec:1}

Microquasars are X-ray binaries with relativistic jets. The microquasar LS 5039 turned out to be the first
high-energy gamma-ray microquasar candidate due to its likely association with the EGRET source 3EG~J1824$-$1514
(\cite{paredes00}). Further theoretical studies supported this association  (\cite{bosch04}), which could be
extended to other EGRET sources (\cite{kaufman02}, \cite{romero03},  \cite{bosch05a}). Very recently,
\cite{aha05} have communicated the detection of the microquasar LS~5039 at TeV energies. This fact confirms the
EGRET source association and leaves no doubt about the gamma-ray emitting nature of this object. The aim of the
present work is to show that, applying a cold-matter dominated jet model to LS~5039, we can reproduce many of the
spectral and variability features observed in this source. Jet physics is explored, and some physical quantities
are estimated as a by-product of the performed modeling. Although at the moment only  LS~5039 has been detected
on the entire electromagnetic spectrum, it does not seem unlikely that other microquasars will show similar
spectral properties. Therefore, an in-depth study of the first gamma-ray microquasar, on theoretical grounds
supported by observations, can render a useful knowledge applicable elsewhere.

\section{A cold matter dominated jet model applied to LS~5039}
\label{sec:2}

The jet is modeled as dynamically dominated by cold protons and radiatively dominated by relativistic leptons. It
considers accretion according to the orbital parameters and companion star properties, allowing for a consistent
orbital variability treatment. The magnetic field energy density and the non-thermal particle maximum energy
values along the jet depend on the cold matter energy density and the particle acceleration/energy loss balance,
respectively, and the amount of relativistic particles within the jet is restricted by the efficiency of the
shock to transfer energy to them. The model takes into account the external and internal photon and matter
fields, as well as the jet magnetic field, all of them interacting with the jet relativistic particles and
producing emission from radio to very high energies. For further details concerning the model, see 
\cite{bosch05b}.

The scenario for LS~5039 is that of a high-mass microquasar. It is an eccentric system (e=0.35) with an O7 star, likely
harboring a black-hole (\cite{casares05}). Concerning the jet geometry, we have adopted typical values present in the
literature. Regarding other jet properties,  for the jet to accretion rate ratio, we have adopted 0.1 and, for the
magnetic field, we have fixed it to 0.1 the equipartition value. To explain observations,  the shock energy dissipation
efficiency the acceleration efficiency have been adjusted to reasonable suitable values ($\sim$10\% of kinetic energy
radiated and efficient acceleration with dE/dt=0.1qBc), similar to those obtained for other microquasars, SNRs, GRBs
and AGNs. The contributions from the disk and the corona have been taken faint according to observational data.

\begin{figure}
\centering
\includegraphics[height=3.1cm,width=6cm]{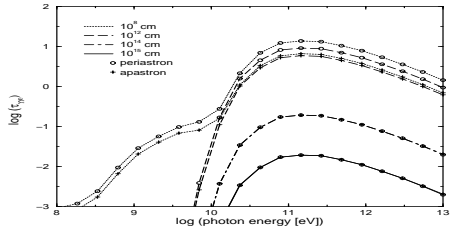}
%
%
\caption{Photon-photon opacity at different jet distances.}
\label{fig:1}       
\end{figure}

\begin{figure}
\centering
\includegraphics[height=5cm,width=7cm]{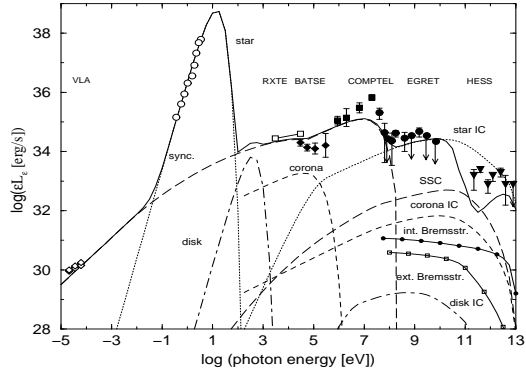}
%
%
\caption{Computed spectral energy distribution of LS 5039.}
\label{fig:2}       
\end{figure}

\begin{figure}
\centering
\includegraphics[height=4cm,width=5cm]{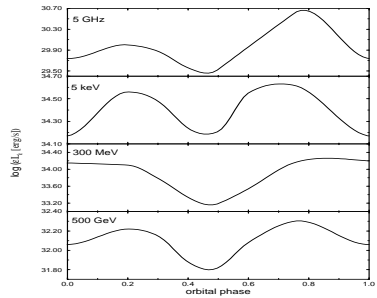}
%
%
\caption{Computed lightcurves of radio, X-ray and gamma-ray radiation.}
\label{fig:3}       
\end{figure}

In Fig.~\ref{fig:1}, we show photon-photon opacities due to the external photon fields for different distances from the
compact object. Although this calculation does not take into account the angular dependence of photon-photon
absorption, it shows clearly that absorption will affect the higher energy band of the spectrum. The computed SED
for LS 5039 is presented in Fig.~\ref{fig:2}. In general, there is a good agreement between the model and the data,
although the model predicts a spectrum in radio that is slightly too hard (see \cite{marti98}). Also, in the TeV
band the fluxes are slightly underpredicted. Both facts can be interpreted as hints of more intense high energy
processes (particle acceleration and emission) taking place outside the binary system, since in our model
radiation is emitted mainly within the 100 GeV-photon absorption region within the binary system. It is worth
noting that the total jet kinetic power required for producing a SED like the one plotted in Fig.~\ref{fig:2} is  pretty
reasonable from the energetic point of view, $\sim 2\times 10^{36}$~erg~s$^{-1}$. In Fig.~\ref{fig:3}, the lightcurve of one orbital
period at different energy bands is presented. Variability has been modeled by orbital motion of the compact
object through a slow asymmetric equatorial wind from a fast rotating stellar companion  (\cite{casares05}). The
slow equatorial asymmetric wind can reproduce a peak at X-rays at phase 0.2, as well as the major peak at phase
0.8 (\cite{bosch05c}). Moreover, at TeV energies there is a peak also around phase 0.8, which resembles what
might be marginally present in HESS data (\cite{casares05}, \cite{aha05}). For more details concerning the
application of the model to LS~5039, see ref. \cite{paredes05}.

%
%
%
%
%

%
%



\printindex
\end{document}